\documentclass[aps,prr,twocolumn]{revtex4-2}
\usepackage{graphicx}
\usepackage{lipsum} % ダミーテキスト
\usepackage{amsmath}
\usepackage{amssymb}
\usepackage{braket}
\usepackage{verbatim}
\usepackage{comment}

\usepackage[section]{placeins}

\begin{document}

\title{Transition between weak and strong measurements in the presence of post-selection}

\author{Shogo Hanashiro}
\author{Holger F. Hofmann}
\affiliation{Graduate School of Advanced Science and Engineering, Hiroshima University,
Kagamiyama 1-3-1, Higashi Hiroshima 739-8530, Japan}

\begin{abstract}
In the weak measurement regime, post-selection can result in the observation of anomalous weak values, seemingly contradicting the eigenvalue statistics observed when the measurement interaction is strong. Here, we investigate the dependence of meter statistics on measurement strength in a post-selected measurement. We find that the meter statistics in the intermediate regime between weak and strong measurements is nearly independent of measurement strength and show that, in this regime, the system performs a measurement of momentum on the meter. Our results show that post-selection is extremely sensitive to the action of the meter on the system, highlighting the difference between the weak measurement regime where this back-action is negligible and the average meter position is a valid measurement result, and the intermediate regime where the meter statistics should not be misinterpreted as a measurement outcome. The transition between weak values and eigenvalues is facilitated by a reversal of the roles of the system and the meter, where the post-selection acts as a readout of information about the meter. The intermediate measurement regime thus reveals that the apparent contradictions between anomalous weak values and eigenvalues originate from the non-classical effects of back-action on the post-selected meter statistics.
\end{abstract}

\maketitle

\section{Introduction}

The role of measurements in quantum mechanics is far from obvious. Observable quantities are represented by operators, and the eigenvalues of these operators can be observed in suitable measurements. However, such measurements must first undo the coherence of superpositions between different eigenstates, requiring a sufficiently strong interaction to achieve the necessary decoherence. The first theoretical analysis of measurement interactions introduced by von Neumann showed that strong measurements initially entangle the system with the meter, producing the expected correlations between the statistics of the meter readout and the eigenstates of the system \cite{Neu}. Since strong measurements are precise, it was long assumed that weak measurements would not add any new insights to von Neumann's measurement theory. In 1988, Aharonov, Albert and Vaidman proved this assumption wrong by showing that the outcomes of post-selected weak measurements could exceed the limits imposed by eigenvalues \cite{AAV}. Specifically, the experimentally observed average meter position is given by the weak value, defined in a time-symmetric manner by both the initial state $\ket{\psi}$ and the post-selected outcome $\ket{f}$ of a subsequent measurement. 

Although weak values were soon confirmed experimentally \cite{exp1}, the implications for measurement theory were initially dismissed and continue to remain controversial \cite{Vai2017}. At the heart of the controversy is the role of post-selection. It has even been pointed out that post-selection can create anomalous measurement outcomes in classical situations, simply by selecting extreme fluctuations of a meter input \cite{FC}. However, such criticism overlooks the obvious fact that weak values are independent of the specific meter setup - if weak values were merely statistical artefacts of a post-selected measurement, different measurement setups should give different results. Instead, weak values provide an objective description of quantum statistics, where weak values of projection operators can be interpreted as Kirkwood-Dirac quasi-probabilities \cite{Lun2011,Lun2012,Hof2012,Sal2013,The2016}. It might be worth noting that Dirac already introduced the definition of what would later be called weak values in 1945, identifying them as the value of an observable for the phase space point defined by an element of the Kirkwood-Dirac distribution \cite{Dir1945}. Dirac's discussion shows that weak values are the quantum analog of physical properties as phase space functions. As such, weak values can describe the relation between incompatible measurements in a manner that identifies the origins of paradoxical correlations with anomalous weak values and the associated negative quasi-probabilities \cite{Aha2002,Wil2008,Yok2009,Den2014,Hof2015,Hig2015,Hal2016,Hof2020,Liu2026}. Since the combination of weak measurement and post-selection refers to the unchanged initial state of the system, the dynamics of the measurement can be neglected. This explains the universality of weak measurements - if the interaction is sufficiently weak, its effects can indeed be neglected and the statistics of the meter have no effect on the measurement outcomes. 

We should now address the elephant in the room. If weak values are objective, how can they depend on the type of measurement used in the post-selection? Is it possible that each post-selection ``creates'' its own reality? This is the core of the measurement problem in quantum mechanics - there is no viable description of a measurement independent reality \cite{Dre2010,Mat2021}. In the weak measurement regime, post-selection depends on the initial coherence of the quantum state and represents a part of the statistics of that state. In the strong measurement regime, the initial coherence is lost and has no effect on future measurements. Aharonov, Bergmann and Lebowitz originally introduced post-selection for a sequence of strong measurements in \cite{ABL}, but it should be obvious that the decoherence caused by a strong measurement changes the post-selection probability. The comparison between the two extreme cases of weak and strong measurements shows that post-selection can work in very different ways. How are these two extreme cases connected? In this paper, we take a closer look at the transition between these two cases in order to identify the relation between the decoherence effects changing the post-selection probability and the information about the objective value of the physical property obtained in the meter readout. Our results show that the onset of decoherence reverses the direction of the information transfer, where the meter statistics is changed by the post-selection, effectively erasing any information about the objective value of the physical property in the process. We can therefore explain how weak values can both be objective and depend on the post-selected outcome. Long before the measurement interaction resolves eigenvalues, the measurement back-action changes the meaning of post-selection, erasing the measurement information transferred to the meter until the back-action is strong enough to completely eliminate the dependence of post-selection on the initial state. 

It is interesting to note that the previous studies of the transition between weak and strong measurements have largely ignored this important aspect of the statistics of post-selection \cite{Pan2020,Ara2021,Tur2023,Son2024,Yua2026,Lima}. Even in \cite{Stammer}, where the problem was described in terms of meter state distinguishability, it was implicitly assumed that the conditional average of meter position can always be interpreted as a physical meter shift. It is precisely this oversight that was criticized in \cite{FC}. Post-selection may cause statistical artifacts, where the
average meter position might originate from initial meter fluctuations selected by their effects on the system. A thorough investigation is needed to understand how the meter statistics are shaped in a post-selected measurement. This is not only a question of our fundamental understanding of quantum mechanics, it also concerns applications in quantum optics, where post-selection is used to shape the quantum statistics of the pointer state \cite{Yua2026}. In the following, we address this problem by analyzing the changes to the meter readout using the relative entropy of the readout distribution and the original Gaussian distribution of the meter state. In this analysis, a meter shift appears as a quadratic increase of relative entropy with measurement strength. Such a quadratic increase is indeed characteristic of both the weak measurement and the strong measurement regime. However, the behavior in the intermediate region is quite different. In the case of anomalous weak values, the rapid increase in relative entropy with measurement strength quickly saturates at a plateau with a measurement strength independent constant value, indicating a ``freeze'' in the post-selected meter shift. This ``freeze'' only ends when the measurement strength starts to resolve the eigenvalues of the system. We show that, in the intermediate regime, the post-selection probability increases with measurement strength while the meter distribution is nearly constant. This observation is explained by the dependence of post-selection probability on the back action of the meter acting on the system. At intermediate coupling strength, the post-selection selects high back action represented by high absolute values of meter momentum. Effectively, the post-selection measures the meter momentum, selecting only the highest values. These are correlated with extremely high meter fluctuations, so that the meter distribution looks nearly identical for a wide range of measurement strengths. 

At intermediate measurement strength, the meter distribution does not represent a measurement of the system. Instead, it is a statistical artefact of the type described in \cite{FC}, where the post-selection identifies extreme fluctuations of the meter input. Importantly, this criticism does not apply in the weak measurement regime or in the strong measurement regime, both of which are legitimate conditional measurements of the observable $\hat{A}$. At intermediate measurement strengths, the increase in post-selection probability and the change in the momentum statistics of the meter indicate that the post-selection process is sensitive to the fluctuations of the initial meter state and modifies the meter statistics by a Bayesian update, not by a meter shift. In the weak measurement regime, the post-selection probability does not depend on measurement strength and no meter information is obtained in the post-selection. The same is true for strong measurement, where the post-selection probability saturates at the value obtained under full decoherence of the eigenstates. The intermediate regime connecting weak and strong measurements is therefore qualitatively different from both, representing a reversal of the measurement direction from a system induced change of the meter to a meter induced change of the system. The scenario analysed in the following illustrates the pitfalls of post-selection in a most striking manner, by showing how Bayesian updates of the meter can effectively wipe out the information transferred from the system to the meter by the measurement interaction. Different from the criticism in \cite{FC}, we can identify the physics of the measurement process, confirming that the problem only arises when measurement strength is in the intermediate regime. Measurements are physical processes, and a detailed understanding of the respective roles of statistics and dynamics is necessary to avoid misunderstandings. 

The remainder of the paper is organized as follows. In section \ref{sec:post}, we briefly review post-selected measurements and distinguish between the weak and strong measurement regimes. In section \ref{sec:stats}, we define the combination of initial state and post-selection in terms of the weak value $w$ obtained in the weak measurement limit and derive analystical expressions for the measurement strength dependent meter distributions. In section \ref{sec:entropy}, we introduce the relative entropy to quantify the changes in the meter distribution caused by measurement interaction and post-selection. The intermediate region is identified as the region where the relative entropy is approximately independent of measurement strength. In section \ref{sec:physics}, the physics behind the meter statistics of the intermediate regime are discussed and the post-selection is identified as a measurement of meter momentum. Section \ref{sec:concl} concludes the paper.

\section{Post-selected measurement}
\label{sec:post}

In classical physics, the state of a physical system is given by a set of physical quantities. All physical properties can be expressed as functions of these physical quantities, and measurements simply map these quantities to a meter system. However, quantum mechanics allows no such description of the measurement process. The initial state cannot be described as a set of quantities, the interaction process entangles the system and the meter, and the readout of the meter system depends on quantum correlations generated by the system-meter interaction. In the strong measurement limit, the initial state $\ket{\psi}$ can be expressed as a superposition of eigenstates $\ket{a}$ of the observable $\hat{A}$ with eigenvalues of $A_a$. The probability of obtaining a measurement outcome $A=A_a$ is then given by $P(A=A_a)=|\braket{a|\psi}|^2$. It is tempting to misunderstand this measurement theory as a description of internal realities, since it completely omits the role of the meter. A proper description of the measurement process starts with a meter system initialized in the state $\ket{\phi}$, so that the total state of system and meter is given by the product state $\ket{\psi}\ket{\phi}$. The measurement interaction will then shift the meter by a value proportional to $\hat{A}$. This interaction is described by the unitary transformation
\begin{equation}
\label{eq:unitary}
    \hat{U}=e^{-ig\hat{A}\hat{p}},
\end{equation}
where $g$ is the measurement strength and $\hat{p}$ is the momentum operator of the meter system. This interaction entangles the system and the meter,
\begin{equation}
    \hat{U}\ket{\psi}\ket{\phi}=\sum_a \braket{a|\psi}\ket{a}\ket{\phi(a)},
\end{equation}
where $\braket{x|\phi(a)}=\phi(x-gA_a)$. After the interaction, the probability distribution of the meter system $P(x)$ is
\begin{align}
    P(x)&=\bra{\psi}\bra{\phi}\hat{U}^{\dagger}\ket{x}\bra{x}\hat{U}\ket{\psi}\ket{\phi} \nonumber \\
    &=\sum_a|\braket{a|\psi}|^2|\phi(x-gA_a)|^2. \label{P}
\end{align}
This probability distribution is easily explained by the eigenvalues $A_a$ and their probabilities $|\braket{a|\psi}|^2$. However, a precise correlation between the meter readouts $x$ and a single eigenvalue $A_a$ is only obtained when the meter wavefunctions $\braket{x|\phi(a)}$ are sufficiently separated from each other. This requires a measurement strength of $g > \sigma$, where $\sigma$ is the initial uncertainty of meter position $\hat{x}$. It is then possible to determine the eigenvalue of the observable $\hat{A}$ from the meter position $x$ obtained in the readout, leaving the system in the corresponding eigenstate $\ket{a}$. If the statistics are accumulated, the relative frequency of the meter readout converges to $|\braket{a|\psi}|^2$. 
If the measurement strength is weak ($g \ll \sigma$), the different wavefunctions are hard to distinguish. The average position of the meter is still given by the expectation value $\braket{\psi|\hat{A}|\psi}$. In general, the average meter position is given by 
\begin{align}
    \braket{\hat{x}} = g\braket{\psi|\hat{A}|\psi}.
\end{align}
 Without post-selection, the measurement determines the expectation value of $\hat{A}$ at all measurement strengths $g$. 

If one performs another measurement on the system after the measurement interaction $\hat{U}$, the outcome $\ket{f}$ of this measurement provides additional information on the meter and the system. In a post-selected measurement, the focus is on a specific measurement outcome $\ket{f}$ that can correspond to arbitrarily rare events. As a consequence, the meter statistics can change drastically due to correlations between the system and the meter generated by the interaction. After a successful post-selection of $\ket{f}$, the normalized state of the meter system $\ket{\phi_f}$ is given by
\begin{equation}
\label{eq:meterstate}
    \ket{\phi_f}=\frac{\bra{f}e^{-i g\hat{A}\hat{p}}\ket{\psi}\ket{\phi}}{||\bra{f}e^{-i g\hat{A}\hat{p}}\ket{\psi}\ket{\phi}||}.
\end{equation}
In general, this state can be expressed as a superposition of shifted meter states $\ket{\phi(a)}$. In the strong measurement regime $(g \gg \sigma)$, these shifted meter states can be distinguished in the readout of meter position. The meter readout statistics are then given by 
\begin{equation}
    P(x) = \sum_a p_{\mathrm{ABL}}(a) |\braket{x|\phi(a)}|^2, 
\end{equation}
where the post-selection modifies the statistics of outcomes $a$ to
\begin{equation}
    p_{\mathrm{ABL}}(a)=\frac{|\braket{f|a}|^2 \;|\braket{a|\psi}|^2}{\sum_{a'}|\braket{f|a'}|^2 \; |\braket{a'|\psi}|^2}.
\end{equation}
Since it was originally introduced by Aharonov, Bergmann and Lebowitz in 1964, We refer to this probability as the ABL probability \cite{ABL}. 
The average post-selected meter position can be determined from the average eigenvalue $A_a$ of the ABL-probability. The result reads
\begin{equation}
    \braket{\hat{x}}_{g \gg \sigma} = g \sum_a A_a p_{\mathrm{ABL}}(a).
\end{equation}
In the strong measurement limit, post-selection only modifies the probabilities of the respective eigenvalues. It has no effect on the meter shifts for each individual outcome $a$. 

Post-selection is more problematic in the weak measurement limit ($g \ll \sigma$). Here, Eq. (\ref{eq:meterstate}) describes a superposition of different meter shifts, and the meter distribution is shaped by quantum interference effects between the different eigenstates $\ket{a}$ in the system. It is therefore interesting to find that the weak measurement limit is independent of the meter state and can be given by the real part of the weak value $A_w$,
\begin{equation}
\label{eq:weak}
    \braket{x}_{g\ll \sigma}=g \; \mbox{Re} (A_w).
\end{equation}
The complex weak value $A_w$ is given by
\begin{equation}
    A_w=\frac{\braket{f|\hat{A}|\psi}}{\braket{f|\psi}}. 
    \label{eq:WV}
\end{equation}
In the weak measurement limit, the shape of the meter wavefunction is almost unchanged. It therefore appears as if the weak value described a single physical shift of the meter that did not match any of the eigenvalues of $\hat{A}$. For small post-selection probabilities ($|\braket{f|\psi}|^2 \ll 1$), the weak values can be extremely high, exceeding the maximal eigenvalues by multiples. Such weak values are commonly known as anomalous weak values. 
It is important to note that such anomalous weak values can only be observed when the meter shift associated with the weak value is much smaller than the initial uncertainty of meter position $\sigma$. This means that weak measurements are limited to $g A_w < \sigma$. The transition from weak measurements to intermediate measurement strengths thus occurs particularly early for anomalous weak values. In the following, we will therefore consider the effects of increasing measurement strength on the post-selected measurement of an extremely anomalous weak value.

\section{Statistics of meter position}
\label{sec:stats}

\begin{figure}[t]
%%\raggedright
%\begin{flushleft}
    \includegraphics[width=\columnwidth]{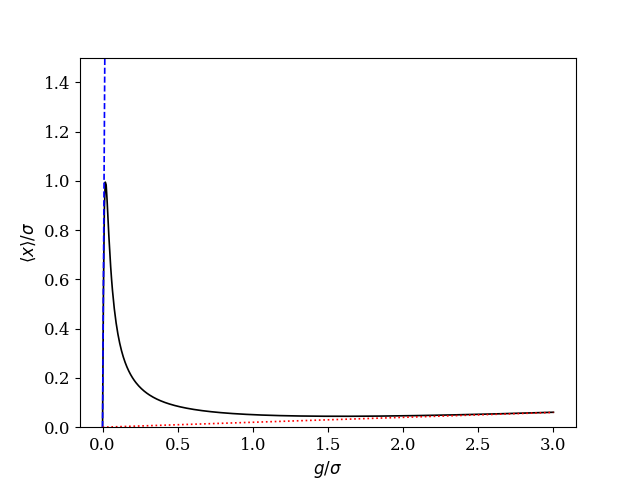} % 段幅に自動調整
%\end{flushleft}
\caption{Average meter position $\braket{x}/\sigma$ after post-selection as a function of measurement strength $g/\sigma$ for an initial weak value of $w=100$. The black solid line shows the complete measurement strength dependence of the average meter position. The blue dashed line illustrates the meter shift in the weak measurement limit $(g \ll \sigma)$, and the red dotted line shows the average meter shift in the strong measurement limit. The slope of the blue dashed line is given by the weak value and the slope of the dotted red line is given by the ABL average.}
\label{fig1}
\end{figure}
In this section, we formulate the transition from weak to strong measurements for a two-level system exhibiting an extremely anomalous weak value. The results on average meter position confirm the observations of the previous section. We will then take a closer look at the actual meter statistics and identify the effects of quantum interference responsible for deviations from eigenvalue statistics. We note that quantum interference effects dominate the statistics well beyond the weak measurement regime. 

We consider the measurement of an observable $\hat{A}\ket{\pm}=\pm\ket{\pm}$ of a two level system, where $\{\ket{+},\ket{-}\}$ is an orthonormal basis of the two dimensional Hilbert space. For simplicity, we choose an equal superposition of the two eigenstates as the initial state of the system,
\begin{equation}
\label{eq:initial}
    \ket{\psi}=\frac{1}{\sqrt{2}}(\ket{+}+\ket{-}).
\end{equation}
For the post-selection, we chose a state with very low probability. We parameterize this state using the real weak value $A_w=w$. This parameterized post-selected outcome is given by
\begin{equation}
    \ket{f}=\frac{1}{\sqrt{2(1+w^{-2})}}\left\{(1+w^{-1})\ket{+}-(1-w^{-1})\ket{-} \right\}. \label{post-selection}
\end{equation}
In the following, we assume that $w \gg 1$. Consequently, $\ket{f}$ is very nearly orthogonal to $\ket{\psi}$ and the post-selection probability at measurement strength zero is then given by
\begin{equation}
\label{eq:weakPf}
    |\braket{f|\psi}|^2 = \frac{w^{-2}}{1+w^{-2}}.
\end{equation}
The post-selection probability is inversely proportional to the square of the weak value. 

In the strong measurement regime, the ABL probabilities are given by
\begin{equation}
\label{eq:pABL}
    p_{\mathrm{ABL}}(\pm) = \frac{1}{2} \pm \frac{w^{-1}}{1+w^{-2}}
\end{equation}
and the average post-selected meter position is
\begin{equation}
\label{eq:strong}
    \braket{\hat{x}}_{g \gg \sigma} = g \frac{2 w^{-1}}{1+w^{-2}}.
\end{equation}
The average meter position increases linearly with measurement strength $g$ in both the weak and the strong measurement regimes. However, the slopes of the increaes are very different, with a high rate of increase for weak measurements and a much slower rate for strong measurement. The transition between the two regimes happens as the overlap between the shifted meter states $\ket{\phi(a)}$ decreases. Here, we consider Gaussian meter states with
\begin{equation}
    \braket{x|\phi}=(2\pi\sigma^2)^{-\frac{1}{4}}\exp\left(-\frac{x^2}{4\sigma^2}\right). \label{meter}
\end{equation}
The uncertainty of the meter state is given by $\sigma$, defining the length scale of the meter shift. Since the eigenvalues are $\pm 1$, the ratio $g/\sigma$ determines whether eigenvalues are resolved or not. Using the Gaussian meter wavefunction, it is possible to determine the post-selected meter distribution $P_{ps}(x)$ at arbitrary measurement strengths $g$,
\begin{align}
    P_{ps}(x) =
    &\frac{1}{\sqrt{2\pi \sigma^2}} \nonumber \\ 
    \times &
    \frac{1}{2((1+w^{-2}) - (1-w^{-2})e^{-\frac{g^2}{2\sigma^2}})}\nonumber \\[0.1cm] \times &\Big((1+2w^{-1}+w^{-2})e^{-\frac{(x-g)^2}{2\sigma^2}}\nonumber \\&-2(1-w^{-2})e^{-\frac{x^2+g^2}{2\sigma^2}}\nonumber\\
    &+(1-2w^{-1}+w^{-2})e^{-\frac{(x+g)^2}{2\sigma^2}}\Big).
    \label{eq:Ppost}
\end{align}
The meter distribution has three distinct contributions, two of which can be identified with Gaussians shifted be the eigenvalues. The third contribution represents quantum interference between the two eigenstates. It is given by a Gaussian centered at $x=0$. The probability distribution in Eq.(\ref{eq:Ppost}) represents the effects of destructive interference between the two eigenstates, hence the interference term must be subtracted from the meter distributions shifted by the eigenvalues. In the actual measurement statistics, quantum interference cannot be separated from eigenvalue contributions, so that the rather simple mathematical formula given by Eq.(\ref{eq:Ppost}) cannot provide a satisfactory explanation of the physics represented by it. 

The post-selected average of meter position is given by
\begin{align}
    \braket{x} &=\int x P_{ps}(x)dx \nonumber \\
    &= g \frac{2gw^{-1}}{1+w^{-2}-(1-w^{-2})e^{-\frac{g^2}{2\sigma^2}}}. \label{eq:x}
\end{align}
We can immediately confirm the strong measurement limit of Eq.(\ref{eq:strong}) by letting the exponential in the denominator go to zero. Likewise, the weak measurement limit of eq. (\ref{eq:weak}) is obtained by setting the exponential in the denominator to one. Eq.(\ref{eq:x}) describes a smooth transition of the average meter position between the weak and the strong regime, similar to the one reported in \cite{Pan2020}. Fig. \ref{fig1} shows the graph of the transition for $w=100$. The weak measurement regime clearly ends as the average meter position $\braket{x}$ approaches a maximal value of $\sigma$, indicating the condition that the meter shift cannot exceed the meter noise in the weak measurement regime. 
\begin{figure}[t]
    \centering
    \includegraphics[width=\columnwidth]{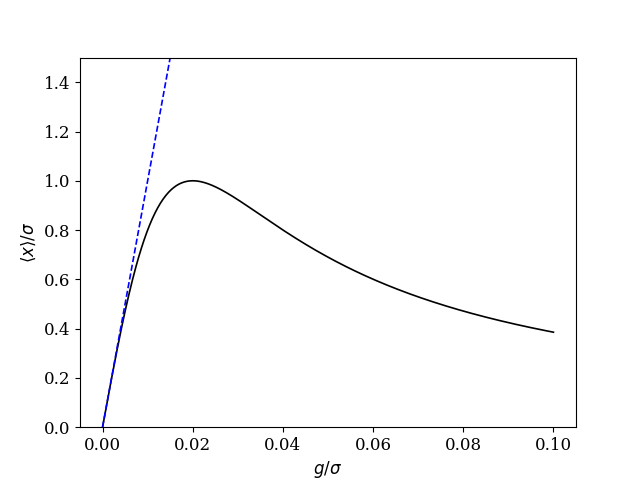}
    \caption{Average meter position $\braket{\hat{x}}/g$ at the transition from the weak measurement regime to the intermediate regime for $w=100$. The average meter position reaches a maxim of $\braket{\hat{x}}/g=1$ at $g/\sigma = 2/w$.}
    \label{fig2}
\end{figure}
Fig. \ref{fig2} shows the transition from the weak measurement regime to the intermediate regime in more detail. In the weak measurement regime $g \ll \sigma$), the post-selected meter distribution $P_p(x,g)$ can be approximated by the derivative in measurement strength at $g=0$,
\begin{equation}
    \frac{dP_p(x,g)}{dg}\Bigg|_{g=0}=\frac{w}{\sqrt{2\pi}} \frac{x}{\sigma^2}  \exp(-\frac{x^2}{2\sigma^2}).
\end{equation}
This derivative describes a meter shift proportional to the weak value,
\begin{equation}
    P_p(x,g)= \frac{1}{\sqrt{2\pi\sigma^2}}\exp{-\frac{(x-gw)^2}{2\sigma^2}}. \label{approximation}
\end{equation}
Thus, the probability distribution in the weak measurement regime can be  approximately by a single Gaussian shifted by $gw$, corresponding to the weak value of $\hat{A}$. The approximation breaks down as $g$ approaches a value of $2 \sigma/w$, where the maximal position average of $\braket{\hat{x}}=\sigma$ is reached. After reaching this maximal value, the average meter position drops only slowly to values much lower than $\sigma$. It is very hard to tell whether this value can still be considered as a measurement result, let alone as a meter shift. Since it actually decreases with measurement strength, the measurement interaction is now changing the meaning of post-selection, and this changed implication of post-selection will be the focus of the following discussion.

\section{Relative entropy of the meter distributions}
\label{sec:entropy}

The average meter position indicates post-selected values of $\hat{A}$ in both the weak measurement regime ($g \ll \sigma$) and the strong measurement regime ($g\gg \sigma$). In the intermediate regime, the relation between average meter position and the values of $\hat{A}$ is less clear. An additional characteristic of the meter distribution described by Eq. (\ref{eq:Ppost}) is needed to keep track of the measurement strength dependence of post-selected meter statistics. Here, we adopt the relative entropy as a measure of the change of the meter statistics. The relative entropy can be interpreted as a statistical distance between two probability distributions. The amount of change in the meter statistics can then be quantified by the relative entropy between the initial position distribution $Q(x)=|\braket{x|\phi}|^2$ and the final meter distribution $P(x)$,
\begin{equation}
    R(P|Q):=\int P(x) \ln\frac{P(x)}{Q(x)}dx.
\end{equation}
The relative entropy quantifies the statistical change of the meter distribution. For a meter shift of $g A_a$ corresponding to the eigenvalue $A_a$ of $\hat{A}$, 
\begin{equation}
    P_a(x)=\frac{1}{\sqrt{2\pi\sigma^2}}\exp\Big(-\frac{(x-g A_a)^2}{2\sigma^2}\Big)
\end{equation}
and 
\begin{equation}
    R(P_a|Q)(g)=\frac{g^2A_a^2}{2\sigma^2}. \label{shifted RE}
\end{equation}
Hence the relative entropy of a meter shift caused by a well-defined value of $\hat{A}$ is quadratic in measurement strength $g$. 

\begin{figure}[t]
\begin{flushright}
    \includegraphics[width=\columnwidth]{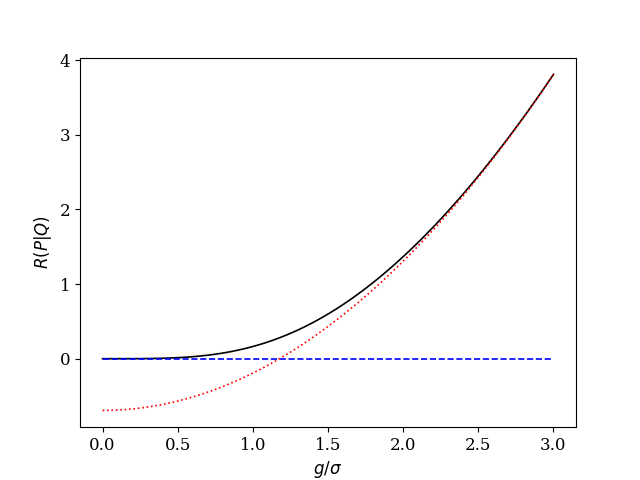}
\end{flushright}
\caption{Measurement strength dependence of relative entropy without post-selection for an initial state with $\braket{\hat{A}}=0$ (black solid line). The blue dashed line indicates the relative entropy of zero observed in the weak measurement limit ($g \ll \sigma$). The red dotted line indicates the strong measurement limit ($g \gg \sigma$). The negative offset of the red dotted line is equal to the Shannon entropy of $H(p_a)=\ln(2)$.}
\label{RE1}
\end{figure}

In the limit of low measurement strength ($g\ll \sigma$), a distribution of shifted Gaussians $P_a(x)$ with probabilities of $p_a$ can be approximated by a single shifted Gaussian distribution of the form
\begin{align}
    P_\Sigma(x) &= \sum_a p_s P_a(x) \nonumber \\ 
    &\approx \frac{1}{\sqrt{2\pi\sigma^2}}\exp\Big(-\frac{(x-g \braket{A})^2}{2\sigma^2}\Big),
\end{align}
where $g\braket{A}$ is the average value of the meter shifts $g A_a$. The slight increase in the variance of the distribution caused by the uncertainty $\Delta A^2$ of the meter shifts can be neglected if $g^2 \Delta A^2 \ll \sigma^2$. At low measurement strength, the relative entropy does not distinguish between a single meter shift and an average meter shift, resulting in the same quadratic dependence of relative entropy on measurement strength,
\begin{equation}
\label{eq:unres}
    R(P_\Sigma|Q)(g) =  \frac{g^2\braket{A}^2}{2 \sigma^2}.
\end{equation}
The situation is different in the strong measurement regime, where the meter shifts are fully resolved ($g\gg \sigma$). When the overlap between the different Gaussian contributions is negligible, the relative entropy is given by 
\begin{equation}
    R(P_\Sigma|Q)(g) \approx \frac{g^2\braket{A^2}}{2\sigma^2}- H(p_a) \label{eq:resolve}
\end{equation}
where $g^2 \braket{A^2}$ is the average value of the squared meter shifts $g^2 A_a^2$ and $H(p_a)$ is the Shannon entropy of the probability distribution $p_a$, 
\begin{equation}
    H(p_a) = \sum_a p_a \ln \left(\frac{1}{p_a}\right).
\end{equation}
The measurement strength dependence in the fully resolved regime is also quadratic, but the rate of increase is increased by the variance $\Delta A^2=\braket{A^2}-\braket{A}^2$. The two approximations intersect at an intermediate measurement strength of 
\begin{equation}
    g_{\mathrm{mid}} = \sqrt{\frac{2 H(p_a)}{\Delta A^2}} \sigma. 
\end{equation}
For classical statistics of meter shifts $g A_a$ given by the probabilities $p_a$, the transition between the weak and the strong measurement regimes is smooth and can be described by a suitable interpolation between the two regimes. This situation applies to measurements without post-selection, where $A_a$ are the eigenvalues of $\hat{A}$ and $p_a=|\braket{a|\psi}|^2$ is the probability of the outcome $\ket{a}$. Fig. \ref{RE1} shows the relative entropy without post-selection for the initial state $\ket{\psi}$ given in Eq.(\ref{eq:initial}). 
In the intermediate regime around $g_{\mathrm{mid}}=1.177 \sigma$, the relative entroly is only slightly higher than the maximum of the two approximations, indicating a smooth transition from unresolved measurements at $g\ll \sigma$ to fully resolved measurements at $g \gg \sigma$. 

The situation changes when we consider the post-selected case. In the weak measurement regime ($g\ll \sigma$), the meter shift is given by the weak value and the relative entropy is approximately given by 
\begin{equation}
\label{eq:weakR}
    R(P_p|Q)(g) \approx \frac{g^2 w^2}{2\sigma^2}.
\end{equation}
In the strong measurement regime ($g\gg \sigma$), ABL statistics apply and
\begin{align}
\label{eq:strongR}
    R(P|Q)(g)&\approx\frac{g^2\braket{A_a^2}_{\mathrm{ABL}}}{2\sigma^2} - H(p_{\mathrm{ABL}}(a)), 
\end{align}
where $\braket{A_a^2}_{\mathrm{ABL}}$ represents the average of the squared eigenvalues $A_a^2$ for the post-selected ABL statistics and $H(p_{\mathrm{ABL}}(a))$ is the corresponding Shannon entropy. For anomalous weak values, $w^2 \gg \braket{A_a^2}_{\mathrm{ABL}}$ ensures that the relative entropy rises much faster in the weak measurement regime than it does in the strong measurement limit. The two approximations never cross and a smooth transition is impossible. 

\begin{figure}
    \centering
    \includegraphics[width=\columnwidth]{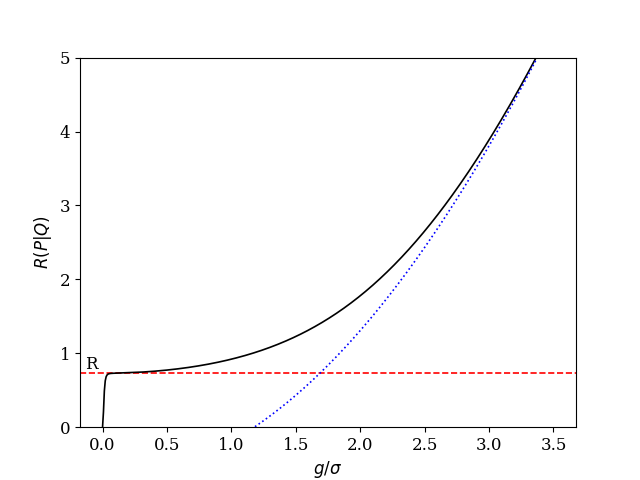}
    \caption{Measurement strength dependence of the relative entropy $R(P|Q)$ in the presence of post-selection for $w=100$. An intermediate region appears between weak and strong measurements, characterized by a plateau of constant relative entropy indicated by the red dashed line. The blue dotted line shows the approximation for the strong measurement regime. 
    \label{plateau1}}
\end{figure}

\begin{figure}[t]
    \centering
    \includegraphics[width=\columnwidth]{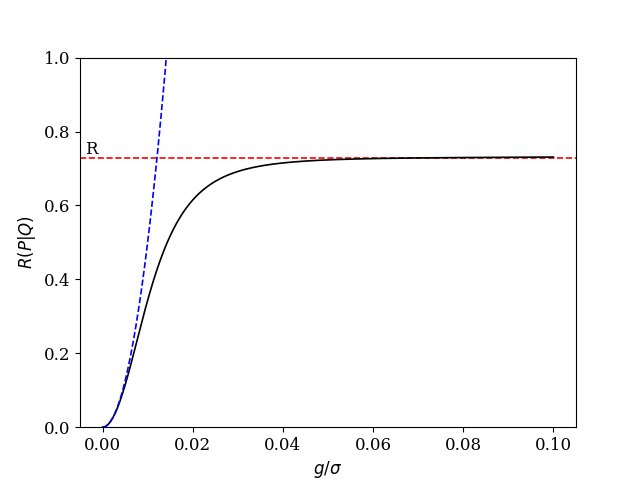}
    \caption{Transition between the weak measurement regime and the intermediate regime. After a rapid quadratic rise, the relative entropy levels off at a plateau value of $R$.The transition happens around $g/\sigma=2/w$.
    \label{plateau2}}
\end{figure}

Fig. \ref{plateau1} shows the relative entropy of the post-selected measurement as a function of measurement strength. The weak measurement regime and the strong measurement regime are connected by an intermediate regime that can be approximated by a plateau at a constant value of $R$. Fig. \ref{plateau2} shows the transition between the weak measurement regime and the intermediate regime in detail. As expected from the average meter shifts shown in Fig. \ref{fig2}, the transition occurs around $g/\sigma=2/w$. The constant value of $R$ indicates that the meter statistics does not depend on measurement strength in the intermediate regime. This means that we are not looking at the effects of a meter shift, since such effects would necessarily continue to increase with increasing measurement strength $g$. In the following, we will take a closer look at the reason for the measurement independent meter statistics in the intermediate regime between weak and strong measurements.

\section{Meter statistics in the intermediate regime}
\label{sec:physics}

The plateau of constant relative entropy in the intermediate regime suggests a third approximation of the meter distribution that cannot be represented by a sum over shifted Gaussian distributions. To identify the appropriate approximation, we first note that the intermediate region exists because the anomalous weak value $w$ is much larger than one, so that the weak measurement condition that requires the meter shift $gw$ to be smaller than the meter variance $\sigma$ breaks down long before the eigenvalues of $\hat{A}$ can be resolved. The intermediate regime can therefore be represented by the limit of $w \to \infty$. Applied to Eq.(\ref{eq:Ppost}), the meter distribution then approaches
\begin{align}
   \lim_{w\to\infty} P_{ps}(x) =
    &\frac{1}{\sqrt{2\pi \sigma^2}}     \frac{1}{2(1-e^{-\frac{g^2}{2\sigma^2}})}\nonumber \\[0.1cm] \times &\Big(e^{-\frac{(x-g)^2}{2\sigma^2}}-2e^{-\frac{x^2+g^2}{2\sigma^2}}+e^{-\frac{(x+g)^2}{2\sigma^2}}\Big).
\end{align}
This limit is sufficient to describe the transition from the intermediate measurement regime to the strong measurement regime. For $g\ll\sigma$, the approximate probability distribution in the limit of $w\to \infty$ can be simplified to 
\begin{equation}
\label{eq:QJstats}
    P_{ps}(x)\approx\frac{1}{\sqrt{2\pi \sigma^2}}\frac{x^2}{\sigma^2}\exp\left(-\frac{x^2}{2\sigma^2}\right).
\end{equation}
This is the approximate meter distribution of the intermediate regime between $g/\sigma=2/w$ and $g/\sigma=1$. The relative entropy $R=R(P_{ps}|Q)$ of this intermediate meter distribution is found to be $R=2-\gamma-\ln 2$, where $\gamma$ is Euler's constant ($\gamma=0.577216...$). The numerical value is roughly $R\approx 0.73$, consistent with the value of the intermediate plateau in Figs. \ref{plateau1} and \ref{plateau2}. 

Using this value, it is possible to identify the precise crossing points of the three approximation of relative entropy. The approximation for the weak measurement regime given by Eq.(\ref{eq:weakR}) achieves a value of $R$ at 
\begin{equation}
    g_{wm}=\sqrt{2(2-\gamma-\ln2)} \; \frac{\sigma}{w}.
\end{equation}
Numerically, this is $g_{wm}\approx 1.21 \sigma/w$, consistent with the crossing point of the approximations shown in Fig. \ref{plateau2}. The approximation for the strong measurement regime given by Eq.(\ref{eq:strongR}) achieves a value of $R$ at
\begin{equation}
    g_{ms}=\sqrt{\frac{2(2-\gamma-\ln(2)+H(p_{\mathrm{ABL}}(a))}{\braket{\hat{A}^2}_{\mathrm{ABL}}}} \; \sigma.
\end{equation}
In general, this crossing point depends on the Shannon entropy and the average $\braket{A_a^2}_{\mathrm{ABL}}$ of the ABL statistics. For the two level system discussed here, $A_a^2$ is equal to one for both eigenvalues, so the ABL average can only be one. In the limit of $w\to \infty$, the Shannon entropy of the ABL probabilities given in Eq. (\ref{eq:pABL}) converges on $\ln(2)$. We can then determine the numerical value of the crossing point of the approximations as $g_{ms}\approx 1.69 \sigma$, consistent with Fig. \ref{plateau1}.

\begin{figure}
    \centering
    \includegraphics[width=\columnwidth]{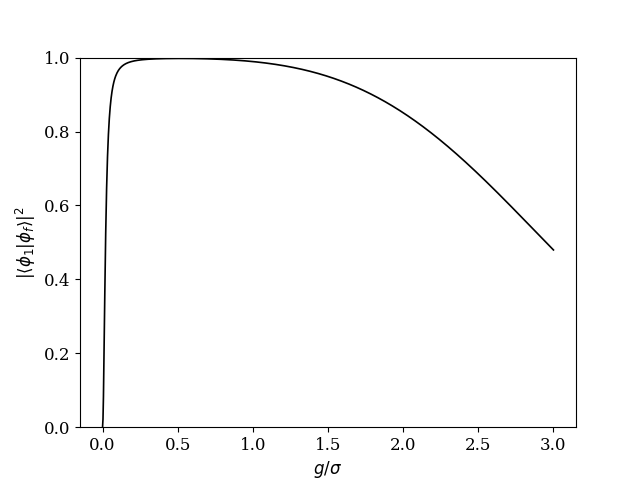}
    \caption{Fidelity $|\braket{\phi_{1}|\phi_f}|^2$ of the first order Gauss-Hermite state as a function of measurement strength.}
    \label{fidelity}
\end{figure}

We have now identified the meter distribution of the intermediate regime, where an increase in measurement strength $g$ has no effect on the meter statistics. 
As the approximation used to derive it shows, the meter distribution given in Eq.(\ref{eq:QJstats}) can be observed experimentally for $w \to \infty$, where the post-selection probability is initially zero and there is no weak measurement regime. This situation has been explored in a very early work of one of the authors, where the input state was a photon vacuum and the post-selected state was a single photon state \cite{Hof2000}. In that work, the meter statistics given by Eq.(\ref{eq:QJstats}) was explained by noting that the quantum fluctuations of $\hat{A}$ observed in the measurement were correlated to virtual photons in the vacuum. Effectively, extreme fluctuations of $\hat{A}$ seemed to be responsible for the observation of $\ket{f}$. The present results indicate a better explanation. The meter readout in this regime should not be interpreted as a measurement result of the system property $\hat{A}$. Instead, the fact that the readout statistics is independent of the measurement strength suggests that the meter state is ``stuck'' in a specific quantum state given by the first order ($n=1$) Gauss-Hermite wavefunction,
\begin{equation}
    \braket{x|\phi_{1}}=\left(2 \pi \sigma^2\right)^{-1/4} \;\frac{x}{\sigma}\; \exp\left(-\frac{x^2}{4 \sigma^2}\right).
\end{equation}
The component of the state $\phi_{1}$ in the post-selected meter state $\ket{\phi_f}$ is given by the fidelity
\begin{equation}
    |\braket{\phi_{1}|\phi_f}|^2 = \frac{\frac{g^2}{2\sigma^2}
    e^{-\frac{g^2}{4 \sigma^2}}}{1+w^{-2}-(1-w^{-2})e^{-\frac{g^2}{2\sigma^2}}}.
\end{equation}
Fig. \ref{fidelity} shows the measurement strength dependence of this fidelity in the intermediate measurement regime for $w=100$. Throughout the intermediate regime, the fidelity $|\braket{\phi_{1}|\phi_f}|^2$ is close to $1$. The meter state conditioned by a post-selection of $\ket{f}$ is thus found to be stuck in a state that is orthogonal to the initial meter state $\ket{\phi}$ ($\braket{\phi|\phi_f} \approx 0$). In the intermediate measurement regime, the post-selection of $f$ selects the changes of the meter state, removing the coherences between the initial meter state $\ket{\phi}$ and the the first order Gauss-Hermite state $\ket{\phi_1}$ that describes the shift in meter position. The only effect of the measurement interaction that is left after the post-selection is a quantum jump to the orthogonal state $\ket{\phi_1}$.  

\begin{figure}[t]
    \centering
    \includegraphics[width=\columnwidth]{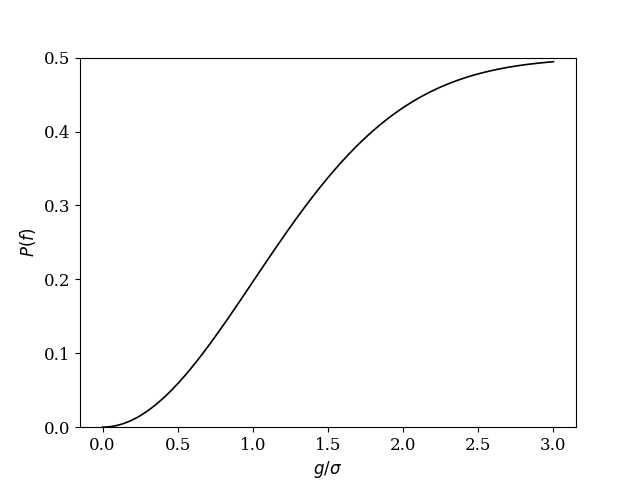}
    \caption{Post-selection probability $P_f$ as a function of measurement strength $g/\sigma$. In the intermediate regime, the post-selection probability increases from its initial value of $|\braket{f|\psi}|^2$ to $1/2$.}
    \label{Pfchange}
\end{figure}

The qualitative change in the meter distribution is related to the low probability of observing $\ket{f}$ in the input state $\ket{\psi}$ of the system. In the limit of $w \to \infty$, the probability of finding $\ket{f}$ in $\ket{\psi}$ drops to zero and the observation of $\ket{f}$ would be impossible without the effects of the measurement interaction. 
The dependence of the post-selection probability $P_f$ on measurement strength $g$ is given by 
\begin{align}
    P_f(g)&=||\bra{f}\hat{U}\ket{\psi}\ket{\phi}||^2 \nonumber \\
    &=\frac{1}{2}\Big(1-\frac{1-w^{-2}}{1+w^{-2}}e^{-\frac{g^2}{2\sigma^2}}\Big).
\end{align}
This equation describes a transition from the initial weak measurement value given by Eq.(\ref{eq:weakPf}) to the strong measurement limit of $P_f=1/2$. Fig. \ref{Pfchange} shows this transition for $w=100$. In the weak measurement regime, the outcome $\ket{f}$ is obtained because it was present in the system before the interaction. For $w=100$, the probability of this outcome is $P_f=10^{-4}$, so it is indistinguishable from zero in Fig. \ref{Pfchange}. For $w\gg1$, the transition from the weak regime to the intermediate regime is approximately described by
\begin{align}
    P_f(g)&\approx w^{-2}\left(1 + \left(\frac{w g}{2 \sigma}\right)^2\right).
\end{align}
The post-selection probability doubles at $g/\sigma=2/w$, at the same point where the average meter position reached its maximal value.
As measurement strength increases beyond this value, the reason for the observation of $\ket{f}$ is the change of the system caused by the measurement interaction, not its initial presence in the state $\ket{\psi}$. In the intermediate regime, the observation of $\ket{f}$ is due to a quantum jump from $\ket{\psi}$ to $\ket{f}$ induced by the measurement dynamics. The meter state $\ket{\phi_1}$ is the signature of this quantum jump on the meter side. Instead of a continuous meter shift, the measurement interaction causes a quantized transition between orthogonal states. 

\begin{figure}
    \centering
    \includegraphics[width=\columnwidth]{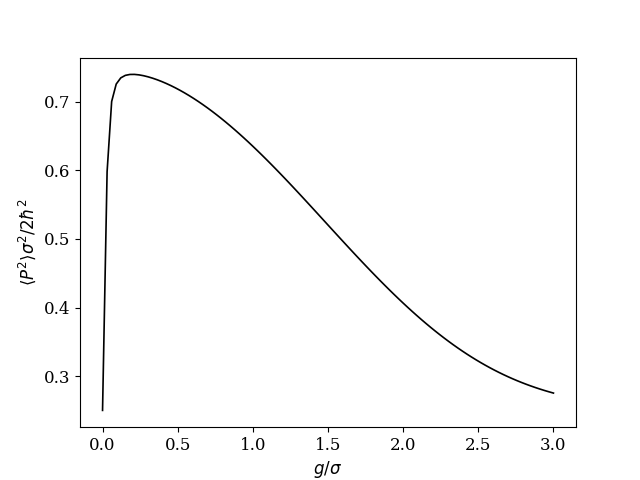}
    \caption{Expectation value $\braket{\hat{p}^2}$ of squared momentum for the post-selected meter state $\ket{\phi_f}$ as a function of measurement strength $g$ for $w=100$. The initial value is equal to the uncertainty of the initial Gaussian, $(\hbar/(2 \sigma))^2$. In the intermediate regime, it rises to about three times this value, corresponding to the uncertainty of $\ket{\phi_1}$. This is followed by a decline back to the original value of $(\hbar/(2 \sigma))^2$.
    \label{momentum1}}
\end{figure}

To understand the physics of this quantum jump, it is necessary to consider the dynamics of the system caused by the meter momentum $\hat{p}$. The interaction given by the unitary transformation in Eq.(\ref{eq:unitary}) changes the system by an amount proportional to the meter momentum $\hat{p}$. This means that the probability of a quantum jump to an orthogonal state of the system is proportional to $\hat{p}^2$. Statistically, the observation of a quantum jump in the system selects components of the meter state with high momentum eigenvalues. Effectively, the post-selection of $\ket{f}$ in the intermediate regime is a measurement of $\hat{p}^2$ in the meter. We can confirm this hypothesis by deriving the expectation value of $\hat{p}^2$ of the post-selected meter state $\ket{\phi_f}$ given by Eq.(\ref{eq:meterstate}). Fig. \ref{momentum1} shows the expectation values $\braket{\hat{p}^2}$ as a function of measurement strength for the transition from weak to strong measurement at $w=100$. In both the weak and strong measurement limits, the post-selected momentum uncertainty is $(\hbar/(2\sigma))^2$, corresponding to the unchanged momentum uncertainty of the initial meter state $\ket{\phi}$. Since the meter momentum is conserved in the measurement interaction, any post-selected increase in $\braket{\hat{p}^2}$ is a result of information obtained about $\hat{p}^2$ from the measurement of $\ket{f}$ in the system. 

The maximal value of $\braket{\hat{p}^2}$ in Fig. \ref{momentum1} is very close to the value of the meter state $\ket{\phi_1}$ dominating the intermediate regime. At sufficiently low measurement strengths, this can be by approximating the post-selected meter state using only $\ket{\phi}$ and $\ket{\phi_1}$. If higher order Gauss-Hermite contributions can be neglected, the expectation value of $\hat{p}^2$ can be approximated by
\begin{equation}
\label{eq:QJ}
    \braket{\hat{p}^2} \approx \frac{\hbar^2}{2\sigma^2}\left(\frac{1}{2} + |\braket{\phi_1|\phi_f}|^2\right).
\end{equation}
As the fidelity of $\ket{\phi_1}$ approaches $1$, the momentum uncertainty approaches three times the value of the initial meter state $\ket{\phi}$. At the low measurement strength end of the intermediate regime, observation of $\ket{f}$ requires values of $\hat{p}^2$ that are three times as high as the initial uncertainty of the meter state. Post-selection statistically selects extreme fluctuations of the meter represented by the state $\ket{\phi_1}$. The approximation of Eq.(\ref{eq:QJ}) above holds as long as the meter system stays within a two-dimensional subspace separating high back action on the system from low back action in a binary manner. The probability $P(f)$ of finding the high back action component of the meter state sufficient to cause a jump to $\ket{f}$ in the system increases with measurement strength.  Eventually, the approximation of Eq.(\ref{eq:QJ}) breaks down as the fluctuations originally present in the meter state are gradually getting strong enough to cause transitions to $\ket{f}$. In the ABL limit, the preference for high momenta vanishes since even small values of $\hat{p}^2$ are sufficient to induce a transition to $\ket{f}$. 

\begin{figure}[t]
    \centering
    \includegraphics[width=\columnwidth]{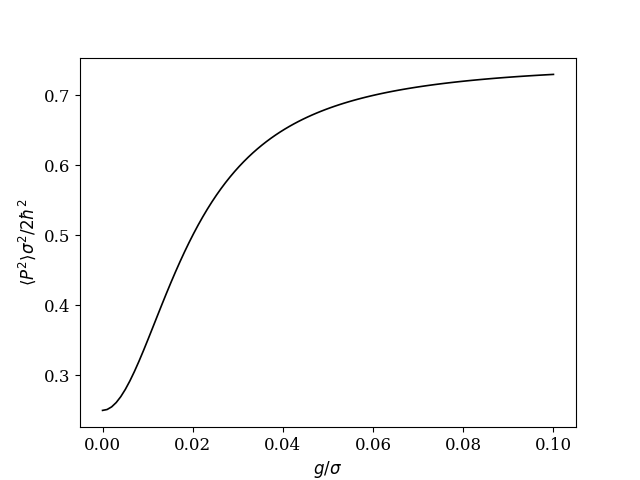}
    \caption{Expectation value $\braket{\hat{p}^2}$ of squared momentum in the transition from the weak measurement regime to the intermediate regime. At $g/\sigma=2/w$, the value is halfway between the uncertainties of $\ket{\phi}$ and $\ket{\phi_1}$.
    \label{momentum2}}
\end{figure}

Fig. \ref{momentum2} shows the transition from the weak measurement regime to the intermediate regime in detail. The approximation given in Eq.(\ref{eq:QJ}) applies and the measurement strength dependence is the same as that of the fidelity $|\braket{\phi_1|\phi_f}|^2$. In the weak measurement regime, the probability of inducing a transition is negligible compared to the initial value of $P_f=1/w^2$. At $g/\sigma=2/w$, the conditional momentum uncertainty is found at the halfway point between the uncertainty of $(\hbar/(2 \sigma))^2$ of the initial meter state $\ket{\phi}$ and the uncertainty $3 (\hbar/(2 \sigma))^2$ of the first order Gauss-Hermite state $\ket{\phi_1}$. Intuitively, this corresponds to an equal split between observations of $\ket{f}$ unchanged by the measurement interaction and transitions to $\ket{f}$ induced by the measurement interaction.  

We can now return to the analysis of meter readout statistics using the relative entropy $R(P|Q)$. The relation between the readout statistics and the momentum statistics is not immediately obvious. In the extreme case of an ideal quantum jump from the Gaussian state $\ket{\phi}$ to the first order Gauss-Hermite $\ket{\phi_1}$, the position and momentum statistics change in the same manner, suggesting a strong positive correlation between $\hat{p}^2$ and $\hat{x}^2$. Since the primary effect of the post-selection is the detection of high momentum values, it follows that the meter readout statistics also represents a statistical update of meter fluctuations that were already present before the measurement interaction. In the intermediate regime, the direction of the measurement is reversed, so that the post-selection of $\ket{f}$ turns into a measurement of $\hat{p}^2$ and, via correlations between meter momentum and meter position, into a statistical update of the readout statistics. It is interesting to note that this corresponds to the concerns expressed in \cite{FC} regarding the conditional average of the meter readout. In that work, it was hypothetically assumed that a hidden mechanism might induce an artificially strong correlation between the post-selected property of the system and the meter position, giving an entirely false impression of the post-selected outcome. Our work shows that this is ruled out by physics in both the weak and the strong measurement regime. We understand the measurement interaction well enough to exclude such hidden processes in the weak measurement regime. However, we have shown here that a more realistic version of the effect described in \cite{FC} does happen in the intermediate measurement regime. Although the conditional average seems to behave in a reasonable manner, the meter readout statistics is determined by correlations between the post-selected outcome $\ket{f}$ and the meter momentum $\hat{p}$ originating from the conditional dynamics of the system. Post-selection thus modifies the fluctuation of initial meter position, making it impossible to identify the contribution of the meter shift in the statistics. The interaction between the system and the meter in a quantum measurement is symmetric in the two systems. It is therefore important to remember that one cannot perform a measurement on the system without also performing a measurement by the system on the meter. In the weak and the strong limit, the information content of this reverse measurement is negligible, so the changes of the meter statistics originate from meter shifts only. In the intermediate regime, this assumption breaks down and the meter statistics originates mostly from the information about the dynamics induced in the system by the meter momentum.

\section{Conclusions}
\label{sec:concl}

The interpretation of post-selected measurement statistics in the intermediate regime between the observation of an anomalous weak value and the fully resolved observation of eigenvalues is a difficult problem due to the inadvertent readout of meter information in the post-selection of the system \cite{FC}. Here, we have identified an intermediate regime characterized by a quantum jump in the meter corresponding to the observation of high momentum fluctuations in the initial meter state. In this case, it is not correct to interpret the statistics of meter positions as a distribution of different meter shifts. 

This insight is particularly important in quantum optics applications, where post-selection of the system is used to shape the quantum statistics of the pointer state \cite{Yua2026}. Our analysis suggests that this kind of operation is best understood in terms of the information about the pointer read out in the post-selection of the system. Quantum interactions are always symmetric, combining any information transfer with a corresponding unitary change of both systems \cite{Mat2021}. Without post-selection, no meter information is obtained and the meter statistics is only modified by the intended meter shift. However, post-selection can provide non-trivial meter information, causing changes to the meter statistics that have no clear relation to the physical property $\hat{A}$. 

The statistical effect that prevents a valid post-selected measurement in the intermediate regime is not observed in either the weak or the strong measurement limit. In the weak limit, the statistics of the meter has no effect on the post-selection, indicating that the criticism of \cite{FC} does not apply and the weak value is objectively valid. The transition to the intermediate regime reveals why it is not a contradiction that objective values of a physical property depend on measurement context. Long before the measurement is strong enough to establish the eigenstates of the physical property as the new measurement context, the back-action of meter momentum completely erases the information that post-selection provided about the physical property of the system. Quantum dynamics ensures that the measurement dependence of reality described by the formalism does not result in any contradictions within our experimentally observable reality. When eigenvalues emerge in the strong measurement limit, the fluctuations of meter momentum have completely randomized the dynamics of the system. Although eigenvalues appear to be objective, it should be noted that they can only be observed after a complete randomization of the dynamics generated by the observable. The difference between eigenvalues and weak values originates from the role of the back-action dynamics in quantum measurements, where eigenvalues are dependent on the disturbance of the system by the back-action induced to observe them \cite{Matsushita}. The intermediate regime connecting weak and strong measurements is dominated by the dynamics of disturbance. Here, the post-selection condition is more likely at extreme meter fluctuations, resulting in the selection of a meter state orthogonal to the initial meter state. This quantum jump characterizes the transition between the observation of anomalous weak values and the observation of eigenvalues in post-selected quantum measurements, demonstrating that the difference between the two regimes originates from the fundamental role of the back action dynamics associated with the meter momentum.

\begin{acknowledgments}
This work was supported by ERATO, Japan Science
and Technology Agency (JPMJER2402).
\end{acknowledgments}

\end{document}